# Development and Clinical Implementation of Next Generation Very Light Weight and Extremely Flexible Receiver Arrays for Pediatric MRI

*Shreyas S. Vasanawala MD/PhD[1], Robert Stormont PhD[3], Scott Lindsay PhD[3], Thomas Grafendorfer MS[3], Joseph Y. Cheng PhD[1], John M. Pauly PhD[2], Greig Scott PhD[2], Jorge X. Guzman MS[3], Victor Taracila PhD[3], Dan Chirayath[3], and Fraser Robb PhD[3]

*Stanford University Depts of Radiology[1] and Electrical Engineering[2], and GE Healthcare[3]*

Correspondence to:

*Shreyas S. Vasanawala

Stanford University
725 Welch Road
Stanford CA 94305
650-724-8087
vasanawala@stanford.edu

**Abstract**

We develop a novel next generation light-weight highly flexible pediatric coil array, combine it with a high-density pediatric posterior array or conventional posterior phased array, and determine feasibility of pediatric clinical use. A highly flexible 16 element MRI receiver coil was constructed with low-profile noise controlling preamplifiers that minimized reactive and resistive coupling. Element decoupling was assessed in flat and highly flexed states. With IRB approval and informed consent/assent, 24 consecutive subjects undergoing torso or extremity MRI were prospectively recruited. Care team members were surveyed on preference for the coil versus conventional coils and diagnostic acceptability of the images was recorded. Confidence interval of proportion of diagnostic exams was calculated. The array without cable weighed 480 grams, demonstrated good flexibility while maintaining element decoupling. The coil was preferred by all nurses and anesthesiologists involved in the care of the patients. Technologists preferred the coil in 96% of cases, and 23/24 exams were diagnostically adequate (ages 23 days – 17 years, mean age 56.7 years, 85% confidence interval of 90-100%). Light-weight highly flexible coil arrays can be constructed that maintain element decoupling. Pediatric clinical image quality is likely to be diagnostic, with acceptance by members of the care team.

**Introduction**

Phased arrays have become ubiquitous in MRI imaging since their original description by Roemer et al. in 1990 [1]. Although the number of coil elements has increased over time the basic nature of the phased array coil has changed little in basic architecture, and has moved from rigid 4 element designs to more flexible architectures, sometimes as high as 64 channels [2-4]. These phased arrays often have element to element isolation of 15dB or less, and such coupling varies widely with the relative proximity and positon of the elements. This has hindered fully capturing the maximal signal to noise benefits of flexible phased array designs in the clinical environment.

More specifically, the pediatric body MRI clinical practice faces challenges of widely varying patient sizes, heterogenous imaging indications, and limited patient cooperation, often necessitating sedation or anesthesia [5-7]. These difficulties are compounded by receiver array coils that are often mismatched to a patient's size [5,8]. In particular, the arrays can be large and thus physically intimidating to young nonsedated patients, limiting their ability or desire to cooperate with the exam. Further, those with abdominal pain experience discomfort with the weight of the array. Additionally, the weight of the coil constricts respiration in sedated children, as well as the youngest patients, even if they are not sedated. This is often managed by raising the coils off the child's chest or abdomen with bolsters or rolled towels to take the weight off the child. Unfortunately, the distress caused by the coils often remains. Further, the performance of the array is degraded by the resulting air gap and lower filling factor, both from a drop in SNR as well as a suboptimal parallel imaging performance.[3]

Semi-flexible coil arrays have been developed to aid patient comfort, enable more tolerance in patient positioning, as well as enhance performance [3,9,10], even at high channel counts [11]. Recently, there has been much movement away from traditional array architectures based upon copper strip arrays towards new technologies including printed conductive inks [4,12,13], along with suitable element decoupling approaches [14,15], to offer a route to even lighter and more form-fitting receiver arrays.

In this work we describe development and clinical acceptance testing of a next generation phased array architecture which facilitates (i) almost 3D complete mechanical flexibility, (ii) improved signal with depth, (iii) an inherent immunity with element / element positioning and (iv) a dramatic reduction in coil weight. Such a phased array architecture is particularly well suited to the field of pediatric MRI. Thus, we combine it with a high-density pediatric posterior array, or for larger patients, a conventional posterior phased array, and determine feasibility of clinical use in a pediatric setting.

**Materials and Methods**

*Coil development:* A coil array was developed with a geometry consisting of sixteen 11cm diameter loops (Fig. 1a), supporting a field of view of 33cm x 37cm. Loops were constructed of a unique metal and dielectric multiple resonator conductor. A custom trace topology is also developed to minimize the reactance of the loop and attain good quality factors of the coil elements, while the overall conductor weight minimized. These loops are interfaced with a new noise controlling preamplifier with low physical profile. These supporting electronics were placed at each loop. This combination and arrangement allows much better control of magnetic and electric field loop to loop coupling and allows element placements outside the historic "critical overlap" regime. Importantly, this enables elements to float over one another. The number of common mode traps was increased, but with lower mass components, to reduce large foci of heat generation.

Performance of the flexible elements was determined by assessing self-inductance, resistance, quality factor, and ratio of loaded to unloaded quality factors for a pair of flexible elements, in comparison to a similar configuration pair of conventional elements. Further, quality factor at a range of overlaps for pairs of conventional and flexible coil elements was determined. Additionally, coil correlation matrices were compared.

The individual loops are packaged in materials comprised of biocompatible technical fabrics for low weight and flammability, and are allowed to float relative to one another for ultimate flexibility. Cabling was positioned on the edge of the array so that the weight of the cable is not supported by a patient, but rather falls on the table to the patient's side. Together, this combination of developments is termed AIR technology.

To enable application across a wide spectrum of clinical indications, the resulting ultra-flexible coil array was made compatible with three posterior array coils. At one end of the spectrum, a high density custom pediatric posterior array with small elements was incorporated [3]. For medium sized patients, the posterior 16 elements of a 32 channel cardiac array (GE Healthcare, Waukesha, WI) was integrated. Finally, for larger fields of view and teenagers, a standard adult posterior array from a 34 channel torso array (Neocoil, Pewaukee WI) was used.

As the impact of physical geometry was an open question, element decoupling of the ultra-flexible array was assessed in both a flat and a highly flexed configuration. Similarly, safety testing, including tests for element heating, arcing, flammability, and pinching were completed. For heating, MRI compatible thermometers were placed over various locations of the coil, and then the coil was loaded and subjected continuously to a pulse sequence with heavy gradients and radiofrequency excitations for a period of four hours. Then a pulse sequence with maximum permissible B1 field was used to ensure no sparks or coil damage resulted. The coil configurations were assessed in volunteers to assess image quality against a standard coil.

*Acceptance testing:* All human subjects participation was pre-approved by the Stanford University Institutional Review Board, performed in HIPPA compliant fashion adhering to all relevant regulations. Informed written consent was obtained from all participants. In a normal volunteer, balanced steady state imaging was performed with the flexible coil and then again with 16 channel anterior array from the conventional cardiac coil (TR 3.6ms, TE 1.6 ms, 8 mm slice thickness). Signal to noise ratio was then measured for each element of each coil. With IRB approval and informed consent/assent, 24 consecutive patients ages 17 and under referred for 3T pediatric MRI of the chest, abdomen, pelvis, or extremities were prospectively recruited over a one week period. MR technologists, nurses, and anesthesiologists associated with each case were surveyed to determine preference for the flexible array compared to conventional coils. The routine imaging protocol was completed, and diagnostic adequacy of images assessed. The 85% confidence interval for the proportion of cases with adequate diagnostic image quality was assessed.

**Results**

Table 1 summarizes characteristics of flexible elements. Of note, flexible elements have higher resistance and unloaded quality factor. However, despite similar radii, the self-inductance of the flexible elements is higher, as is loaded quality factor. Further, a greater preservation of quality factor over a range of coil overlaps was noted (Figure 1b, Figure 1c, and Table 1). Additionally, in a two-element configuration, the off-diagonal correlation matrix elements were .695 unloaded and .168 loaded for traditional coil elements, but only .339 unloaded and .143 loaded for flexible elements.

The complete resulting coil array with sixteen elements weighed 480 grams without the cable. The coil demonstrated extreme flexibility, as seen in Figure 2. Phantom testing demonstrated the elements were highly decoupled from each other (<-20db between channels) as seen in Fig 3. Measured element $Q_{unloaded}$ / $Q_{loaded}$ ratios were approximately 4.5. The coil demonstrated good flexibility in all directions (Fig 3). Noise correlation data was collected both with the array flat and flexed around cylinder with no deterioration found (Fig 3).  Temperature assessment immediately after high SAR and gradient sequences demonstrated no concerning foci of heating.  Similarly, remaining safety tests raised no concerns.  A volunteer scan using the flexible coil and a conventional cardiac array of similar size showed equivalent image quality and signal to noise ratio (Fig 4).

Recruited subjects and clinical indications are shown in Table 2, spanning musculoskeletal, abdominal, and cardiovascular exams.   Ages ranged from 23 days to 17 years, with mean of 5.67 years, and ten males.  Eight cases were performed without sedation.  Despite significant physical manipulation, including flexing and folding, as well as placing intravenous pumps and other equipment on the coil, all elements were functional on the final patient, and no repairs were required.

Anesthesiologist and nurse preference were uniformly for the flexible coil (16/16 and 18/18 cases). Technologist preferred the flexible coil in 96% of cases, and each of those cases was diagnostically adequate.  The 85% confidence interval for the proportion of diagnostically adequate cases was 90.2% – 100%.  In one case, coil coverage for long bones of the leg was not adequate. In no case was the coil offset from the patient with bolsters due to concern of the weight of the coil restricting respiration.  Representative images are shown in Figures 5, 6, and 7, which show outstanding image quality.

**Discussion**

Construction of a highly flexible form-fitting coil array with very light weight is feasible. This is enabled in part by integration of miniaturized supporting electronics and locating them at each loop. This co-location assists in element isolation and lowering power losses. Element to element decoupling was found to be highly robust to the coil wrapping or flexing. The coil is likely to be preferred by anesthesiologists and pediatric nurses, and yield diagnostically adequate images. Although not assessed in this work, this coil architecture is likely to benefit non-pediatric applications, such as fetal MRI, where weight and thickness of the coil is an issue, bariatric patients, and perhaps joints of adults.

In our own practice, current commercially available MR coils suffer in a pediatric setting from several issues. First, the weight of the coil is too great for anesthetized patients, compromising a patient's respiration. For the non-anesthetized patient, the weight of the coil is uncomfortable and intimidating. Because of these concerns, the coils are lifted off the patients with a bolster, typically composed of rolled towels. Unfortunately, this results in poorer filling of the coil, lower signal to noise ratio due to distance from the organs of interest, and poorer parallel imaging performance. Second, because the coils have some rigidity, even without the weight limitations, they are not form-fitting. Again, this results in lower parallel imaging performance and lower signal to noise in the images. This work addresses both of these issues by enabling very light weight and highly flexible coils.

There are several limitations to the work. First, although the weight of the coil array itself has been reduced, the cabling is still heavy and bulky. While it has been positioned to the side of the array, and thus should not weigh on the patient, it still can intimidate children. Similarly, packaging material for this coil architecture can be further optimized, using thinner materials, as well as adornments that are appealing to children. Thus, further refinements to the array should be made.

Second, although the coil array performance is assessed quantitatively, the in vivo performance is qualitatively assessed, without a comparison to a conventional coil array. The challenge with in vivo quantitative assessment lies with the standard use of parallel imaging in clinical protocols, which precludes a true SNR measurement. The challenge to obtaining a comparison with a conventional coil is the lack of cooperation in a pediatric unanesthetized population and the precluding safety considerations of prolonging anesthesia, while the coils are swapped and the patient repositioned in the scanner. Finally, the sample size of the study suggests feasibility of use of a flexible array and its preference by pediatric practitioners across a range of applications, but a larger assessment in multiple specific applications is the subject of future work.

**Acknowledgements**

We gratefully acknowledge support of NIH R01EB019241, R01 EB009690, Tashia and John Morgridge Faculty Scholars Fund, and P41EB015891. Additionally, authors thank Piero Ghedin for assistance with volunteer scanning.

**Competing Financial Interests**

Some authors are employees of GE Healthcare. However, the remaining authors, and in particular the corresponding author, designed the studies, had full control of data collection, full control of data analysis, drafting of the manuscript, and review of the final manuscript.

Specifically, authors RS, SL, TG, VT, JXG, DC, and FR are emloyees of GE Healthcare, a company which may stand to gain or lose financially from this publication.

Authors SSV, JYC, GS, JMP have no competing financial interests.

# Figures

Figure 1. Coil performance is maintained over a range of element overlaps. (a) Flexible anterior array (blue) on top of high-density pediatric posterior array (cream). Insets show cabling and individual element with minimal profile feedboard. (b) Arrangement of two elements with centers at a given separation, (c) quality factors (Q) for conventional and flexible technologies show good preservation across a range of overlaps for the flexible technology.

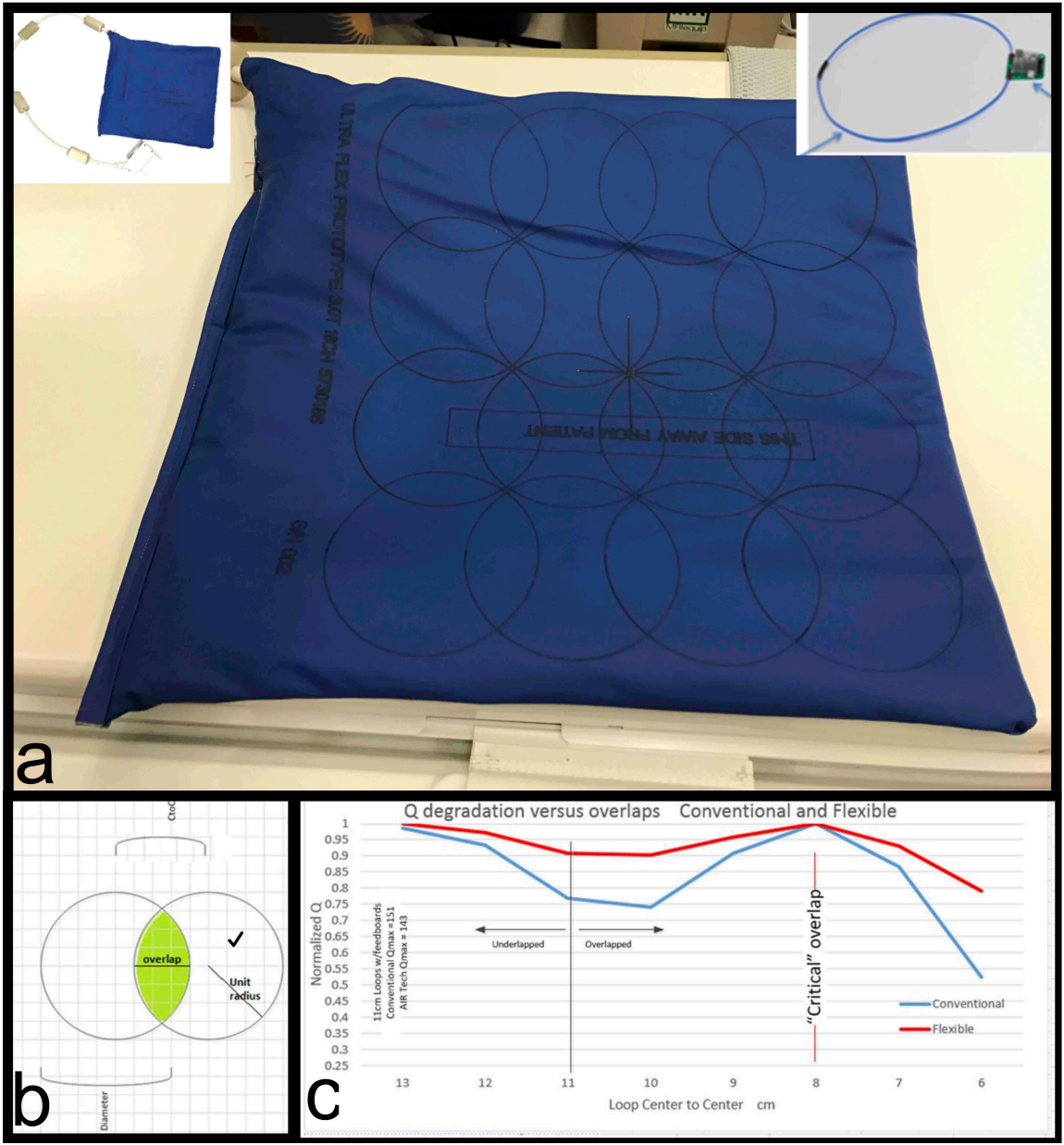

Figure 2. New coil demonstrates unprecedented flexibility. In a flat configuration (bottom) , the coil array sits on a coffee mug. The array can be folded to fit into the cup (top left). Despite this exteme flexibility, the coil array maintains functionality.

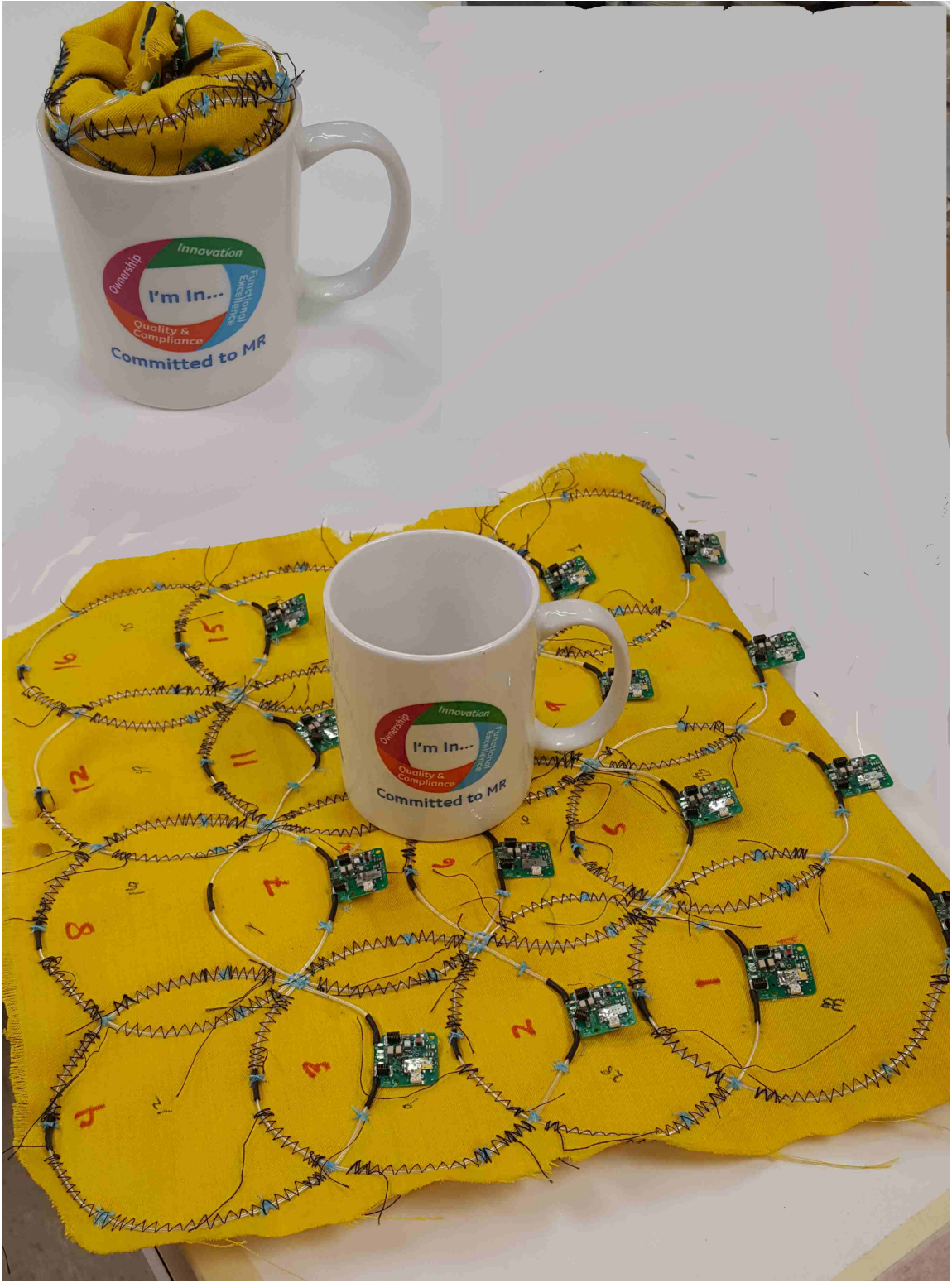

Figure 3. Element decoupling is maintained despite flexing. *Left top*: Individual element images from flexible anterior array of a slab phantom shows excellent decoupling, with a well-defined focus of signal reception under each of the sixteen elements. *Left bottom*: Easy flexibility in right/left and antero-superior directions is noted. *Right*: unloaded noise correlation matrix in flat and right/left flexed configuration to 13 cm diameter shows well preserved element decoupling.

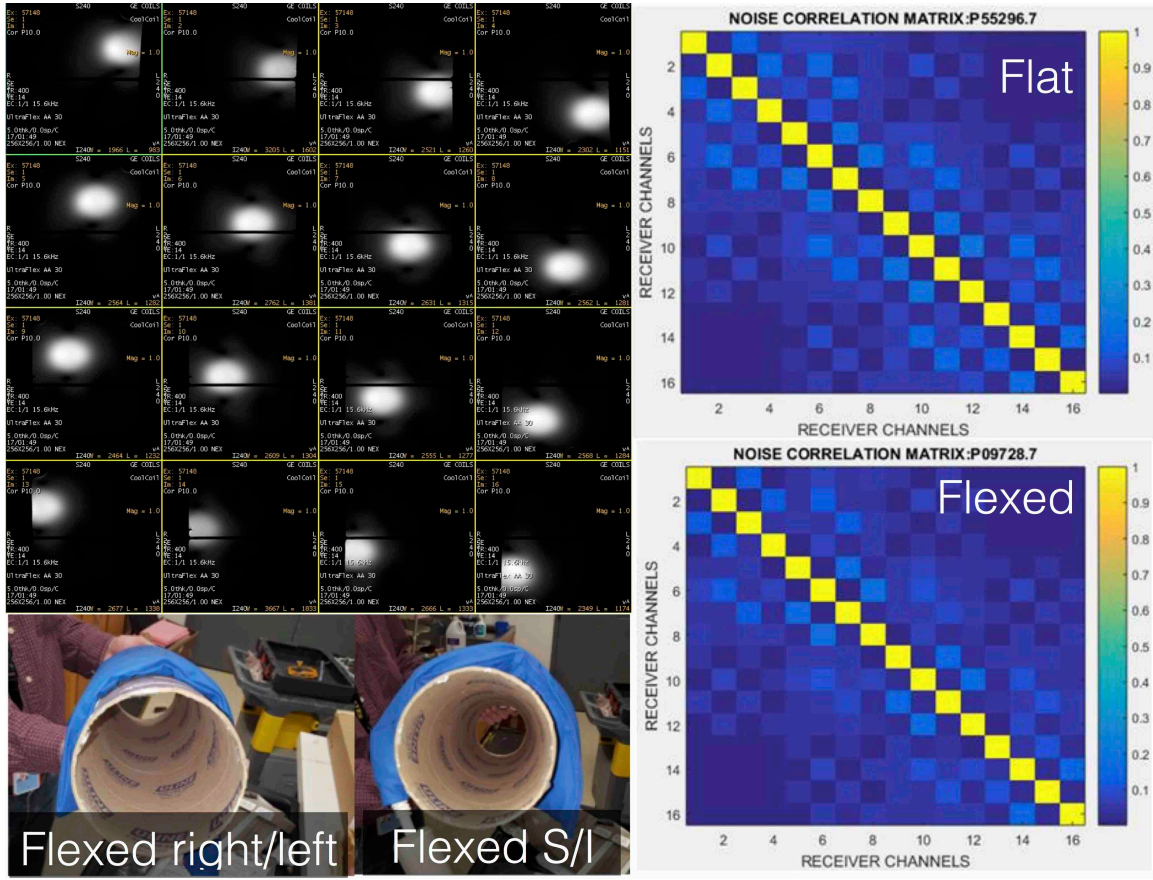

Figure 4. Comparison of image quality in an adult volunteer. A conventional 16 channel coil (a) and the flexible coils (b) yield similar image quality. Additionally, SNR for individual elements (c) in a representative slice shows similar performance.

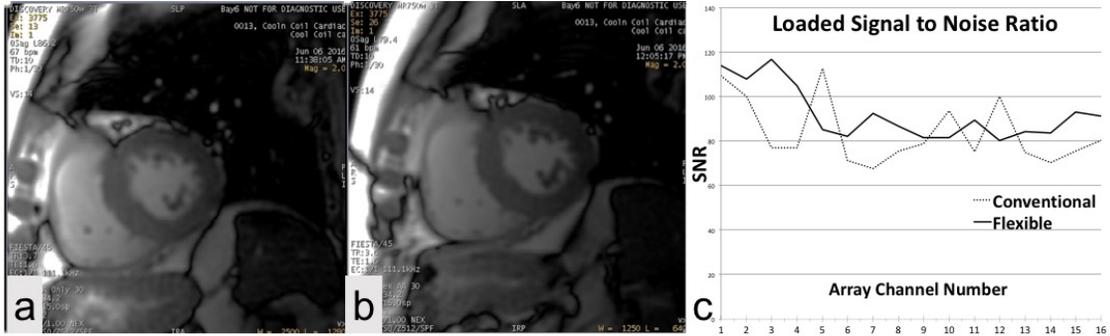

Figure 5. Musculoskeletal MRI with only the flexible coil. Top: Coronal volumetric T2-weighted fast spin echo 1.2 mm thick slices of a child with fixed hip flexion and femur deformity, and coil wrapped around right hip. Bottom: Sagittal T1-weighted imaging of elbow fracture with fixed flexion, also obtained with wrapping of coil. Note that challenges with patient positioning are common in pediatric musculoskeletal MRI, which are well addressed in these two cases.

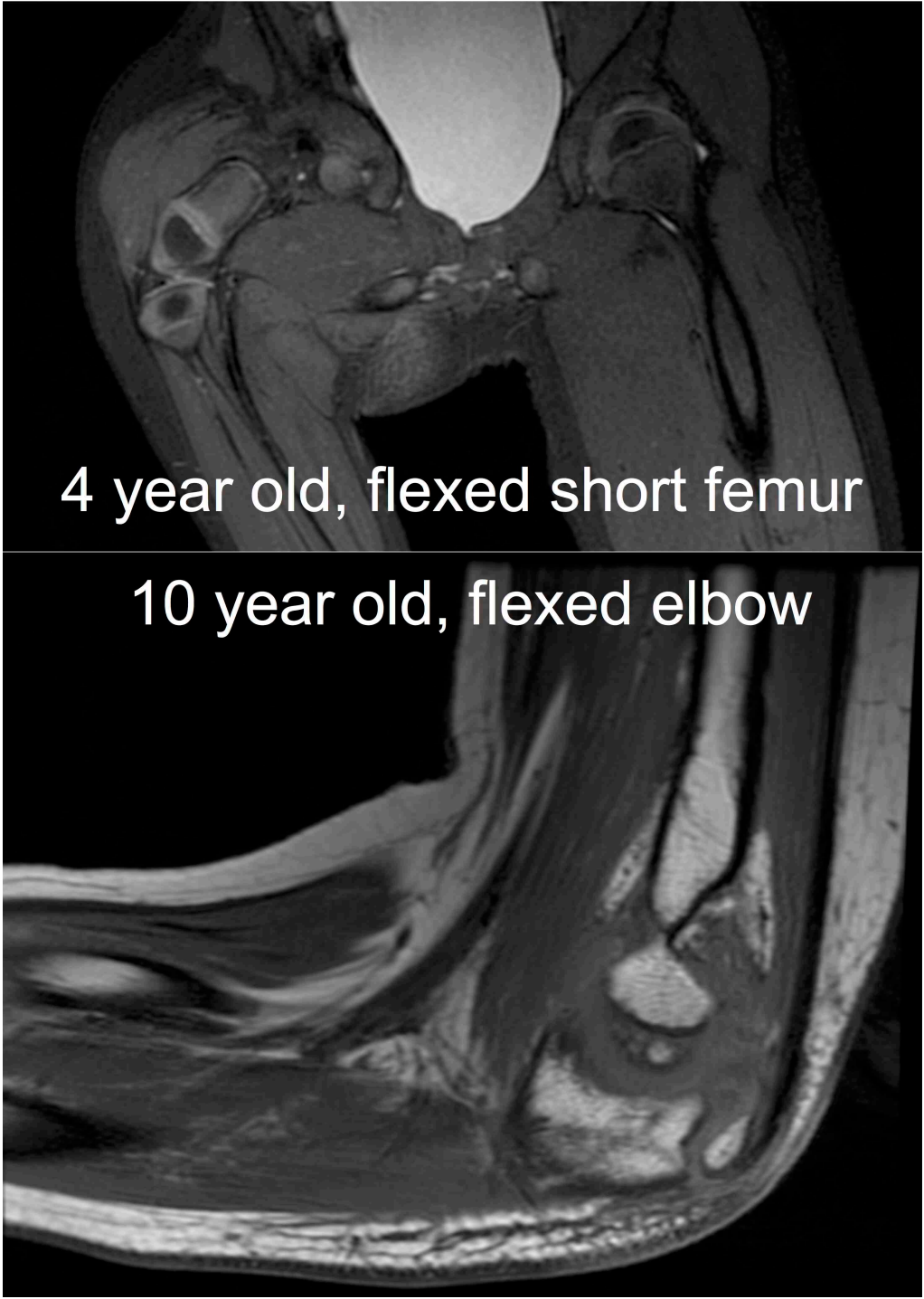

Figure 6. Free-breathing abdominal MRI by combining the light flexible array anteriorly with a high density pediatric posterior array. (top) Coronal fat-suppressed T2-weighted single shot fast spin echo imaging of 1 month old (TE 130 ms, FOV 24 cm), and (bottom) 5 year old (3 mm slice thickness, FOV 28 cm, TE 91 ms). Note excellent subjective SNR despite high spatial resolution and subsecond imaging. (c) Coronal MIP from 3D contrast-enhanced MRA at 2.4 mm slice thickness with 320 x 224 matrix over 22 cm FOV and (d) axial 3D T1-weighted delayed post-contrast imaging at 2 mm slice thickness and 320x320 matrix over 18 cm FOV of a 9 month old with a liver tumor.

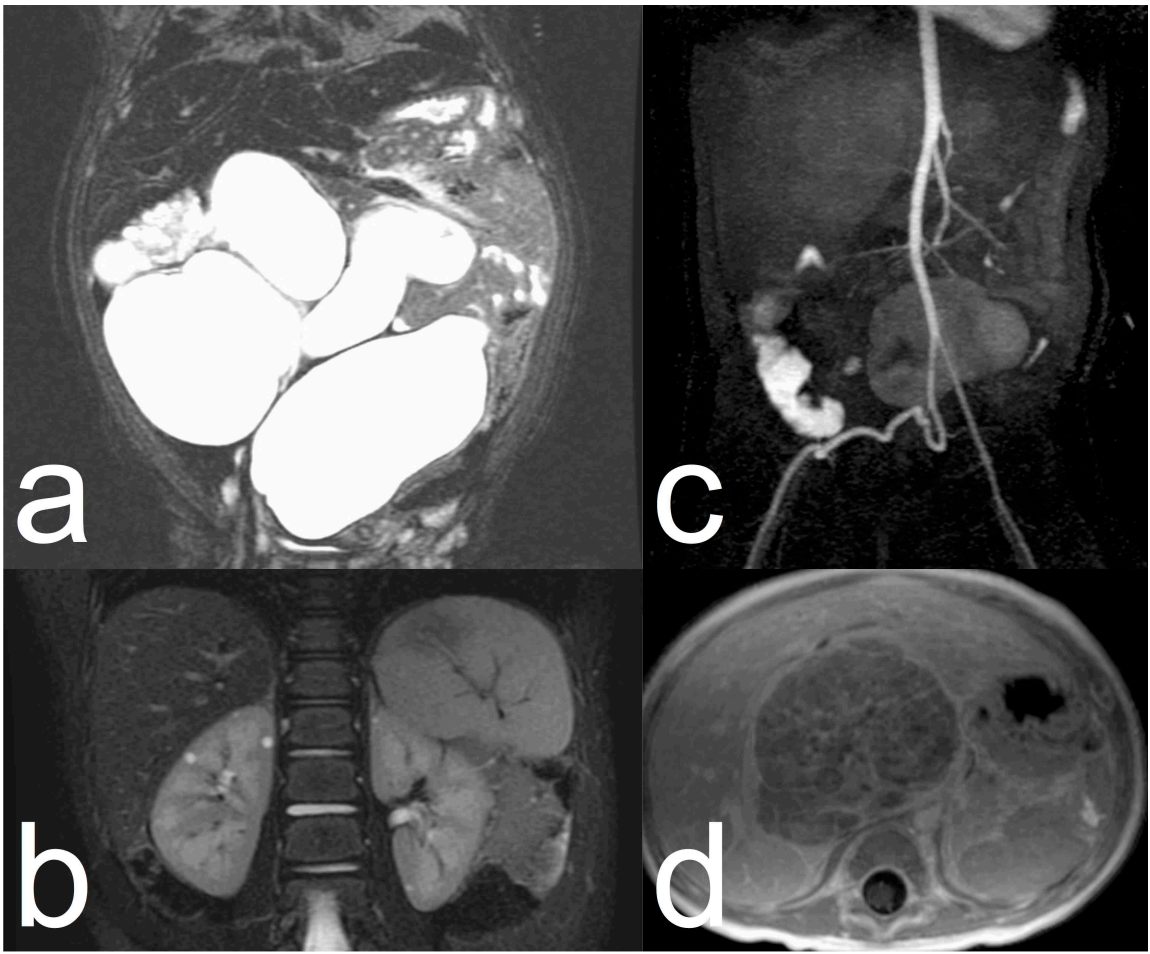

Figure 7. Cardiovascular imaging with flexible anterior array and high density pediatric posterior array. 23 day old non-sedated premature baby weighing only 1.4 kg, with (a) short axis cardiac gated single shot black blood fast-spin echo imaging at heart rate of 131 beats per minute (FOV 20 cm with 256 x 192 matrix, 4 mm slice thickness, echo time 110 ms) revealing noncompaction (white arrow), and (b) ferumoxytol enhanced 4D flow (0.7 mm in plane resolution, 1 mm slice thickness, TR 3.9 ms, TE 1.7 ms, 16 millisecond true temporal resolution, venc 250 cm/s) showing atrial septal defect (dashed white arrow), mitral valve (black arrows), and reformat capability.  Coil was resting directly on the delicate child with no need for offsetting the weight of the coil.  Same array on a 10 year old at heart rate of 94 beats per minute (c), with ferumoxytol-enhanced 4D flow (0.75 mm in-plane resolution, 1.4 mm slice thickness, 21 millisecond true temporal resolution, venc 250 cm/s, TR 4 ms, TE 1.8 ms) showing outstanding tricuspid valve  delineation (black dotted arrow), refomat capability, and diastolic flow in the right coronary artery (dotted white arrow).

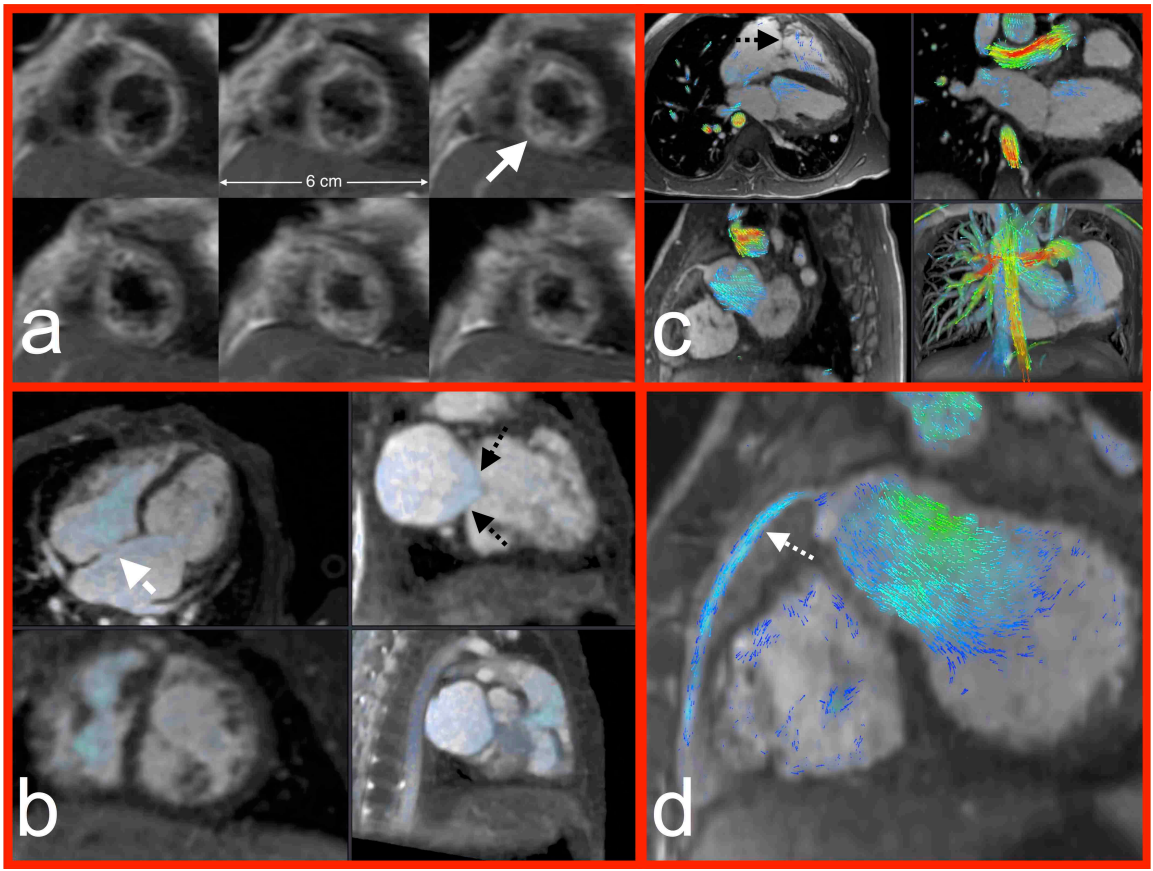

Table 1. Performance characteristics of a pair of conventional coil elements and a pair of flexible coil elements: self-inductance (L), resistance (R), quality factor ($Q = 2\pi f_0 L/R$), ratio of unloaded to loaded quality factors ($R_Q$), and $R_{SNR} = \sqrt{1 - 1/R_Q}$. Last three columns show quality factor for conventional and flexible coils at a range of spacings between the centers of the two elements.

| | L (uH) | R (Ω) | Q | $R_Q$ | $R_{SNR}$ | Space (cm) | Conventional | Flexible |
|---|---|---|---|---|---|---|---|---|
| Conventional unloaded element 1 | .346 | 1.1 | 253 | 21.38 | 97.63 | 6 | 79 | 113 |
| Conventional loaded element 1 | .331 | 22.5 | 11.8 | | | 7 | 131 | 133 |
| Conventional unloaded element 2 | .346 | 1.1 | 253 | 22.84 | 97.97 | 8 | 151 | 143 |
| Conventional loaded element 2 | .331 | 24.0 | 11.1 | | | 9 | 137 | 137 |
| Flexible unloaded element 1 | .532 | 2.4 | 178 | 10.77 | 95.25 | 10 | 112 | 129 |
| Flexible loaded element 1 | .502 | 24.4 | 16.5 | | | 11 | 116 | 130 |
| Flexible unloaded element 2 | .532 | 2.4 | 178 | 10.16 | 94.95 | 12 | 141 | 139 |
| Flexible loaded element 2 | .504 | 23.1 | 17.5 | | | 13 | 149 | 143 |

Table 2. Imaged subjects in different configurations: flexible coil alone (N), with small pediatric posterior array (S), with the posterior 16 elements of a 32 channel adult cardiac array (C), and the posterior elements of a standard large adult posterior array (L). Preference for flexible coil is denoted by a "+", while preference for a conventional coil is denoted by a "-".

| Subject | Indication | Coil | Age | Anesthesia | Tech | Nursing | Exam |
|---|---|---|---|---|---|---|---|
| 1 | Abdominal mass | S | 5 m | N/A | + | + | + |
| 2 | Myocarditis | S | 23 d | N/A | + | + | + |
| 3 | Kidney anomaly | S | 1 m | + | + | + | + |
| 4 | Repaired tetralogy | S | 10 y | + | + | + | + |
| 5 | Elbow trauma | N | 10 y | N/A | + | N/A | + |
| 6 | Osteomyelitis | N | 14 y | N/A | - | N/A | - |
| 7 | Kidney mass | S | 5 y | + | + | + | + |
| 8 | Leg deformity | S | 4 y | + | + | + | + |
| 9 | Liver mass | S | 9 m | + | + | + | + |
| 10 | Abdominal pain | L | 17 | N/A | + | N/A | + |
| 11 | Osteomyelitis | L | 3 y | + | + | + | + |
| 12 | Diffuse liver | L | 1 y | + | + | + | + |
| 13 | Spine& rectal | L | 3 y | + | + | + | + |
| 14 | Casted elbow | L | 13 y | N/A | + | N/A | + |
| 15 | abdominal mass | L | 3 y | + | + | + | + |
| 16 | bone tumor | L | 13 y | N/A | + | N/A | + |
| 17 | kidney mass | L | 5 y | + | + | + | + |
| 18 | aortic coarctation | L | 9 y | N/A | + | N/A | + |
| 19 | pulmonary atresia | C | 3 y | + | + | + | + |
| 20 | coronary anomaly | C | 5 y | + | + | + | + |
| 21 | pulmonary | C | 12 y | + | + | + | + |
| 22 | neuroblastoma | C | 10 m | + | + | + | + |
| 23 | germ cell tumor | C | 3 y | + | + | + | + |
| 24 | forearm nodules/ | C | 4 y | + | + | + | + |
| Total | | | | 16/16 | 23/24 | 18/18 | 23/24 |